\documentclass[twocolumn,english,pra,prl,floatfix,showpacs]{revtex4}
\usepackage[T1]{fontenc}
\usepackage[latin9]{inputenc}

\usepackage{babel}

\begin{document}

\title{Irreducible multiparty correlation can be created by local operations}

\author{D.L. Zhou}

\affiliation{Beijing National Laboratory for Condensed Matter Physics, and Institute
of Physics, Chinese Academy of Sciences, Beijing 100190, China}
\begin{abstract}
Generalizing Amari's work titled {}``Information geometry on hierarchy
of probability distributions'' \cite{Amari}, we define the degrees
of irreducible multiparty correlations in a multiparty quantum state
based on quantum relative entropy. We prove that these definitions
are equivalent to those derived from the maximal von Neaumann entropy
principle \cite{Linden,Zhou}. Based on these definitions, we find
a counterintuitive result on irreducible multiparty correlations:
although the degree of the total correlation in a three-party quantum
state does not increase under local operations, the irreducible three-party
correlation can be created by local operations from a three-party
state with only irreducible two-party correlations. In other words,
even if a three-party state is initially completely determined by
measuring two-party Hermitian operators, the determination of the
state after local operations have to resort to the measurements of
three-party Hermitian operators. 
\end{abstract}


\maketitle

\textit{Introduction.} ---
In quantum mechanics, the complete information of a multiparty system
is contained in its multiparty quantum state, which can be revealed
by performing different types of quantum measurements. A natural classification
of the types of quantum measurements on a multiparty system is based
on the number of parties involved in a measurement. In other words,
a $k$-party measurement in an $n$-party system is represented by
a $k$-party $(1\le k\le n)$ Hermitian operator. By measuring $k$-party
Hermitian operators, we can determine the $k$-party reduced density
matrices. Obviously, the $(k-1)$-party reduced density matrices are
determined by the $k$-party reduced density matrices. Thus the degree
of $k$-party irreducible correlation is defined as how much more
information contained in the $k$-party reduced density matrices but
nonexistent in the $(k-1)$-party reduced density matrices. The concept
of irreducible $n$-party irreducible correlation for an $n$-party
quantum state was first proposed in Ref. \cite{Linden} by Linden
\emph{et al.}, and we generalized it to irreducible $k$-party correlations
for any $1\le k\le n$ in an $n$-party state in Ref. \cite{Zhou}.

As mentioned in Ref. \cite{Zhou}, the irreducible $k$-party correlation
in an $n$-party state can be regarded as the quantum version of the
connected information of order $k$ for a probability distribution
of $n$ classical variables defined in Ref. \cite{Schneidman}by Schneidman
\emph{et al.}. However, in classical information community, there
exists a related work titled {}``Information geometry on hierarchy
of probability distributions'' \cite{Amari} by Amari. Thus the following
questions naturally arise: What is the quantum version of Amari's
work? Is the quantum version equivalent to the irreducible correlations
defined in Refs. \cite{Linden,Zhou}? Does it give us new insight
on irreducible multi-party correlations? Here we remark that, even
in classical information community, as far as we know, one does not
know whether the definitions given in Ref. \cite{Amari} and in Ref.
\cite{Schneidman} are equivalent.

Another related interesting topic on multiparty correlations is to
define and characterize the genuine $n$-party correlation in an $n$-party
quantum state \cite{Zhou1,Winter,Horodecki}. Is the degree of irreducible
$n$-party correlation in an $n$-party state \cite{Linden} a legitimate
measure of genuine $n$-party correlation?

In this paper, we give the quantum version of Amari's work, and we
prove that this quantum version is equivalent to the irreducible correlations
defined in Refs. \cite{Linden,Zhou}. Based on this equivalent form,
we find a counterintuitive result: the irreducible three-party correlation
can be created by local operations from a three-party quantum state
with only irreducible two-party correlations. This implies that local
operations can create higher order irreducible correlations from lower
order correlations. In other words, the degree of irreducible $n$-party
correlation in an $n$-party state can increase under local operations,
and it is not a measure of genuine $n$-party correlation in an $n$-party
state \cite{Zhou1,Winter,Horodecki}.

For the sake of notation simplicty and without losing of generality,
we will formulate our problem for a three-qubit system.

\textit{Two representations of a three-qubit quantum state.} ---
To simplify our presentation, we shall adopt the following notations.
First, let $\mathbf{m}$ denote $(m_{1},m_{2},m_{3})$, with $m_{i}\in\{0,1,2,3\}$
for $i=1,2,3$. Specifically, $\mathbf{0}$ denotes $(0,0,0)$, and
$\bar{\mathbf{m}}$ can take any values as $\mathbf{m}$ except $\mathbf{0}$.
Second, let $\mathbf{\sigma}_{\mathbf{m}}^{(123)}$ denote $\sigma_{m_{1}}^{(1)}\otimes\sigma_{m_{2}}^{(2)}\otimes\sigma_{m_{3}}^{(3)}$,
where $\sigma_{0}$ is the $2\times2$ identity matrix, and $\sigma_{i}\;(i=1,2,3)$
are three Pauli matrices. Third, let $N_{0}(\mathbf{m})=\sum_{i}\delta_{m_{i}0}$,
and $[3]=(123)$.

An arbitrary three-qubit state $\rho^{[3]}$ with maximal rank can
be written in the exponential form \begin{equation}
\rho^{[3]}(\{\theta^{\mathbf{m}}\})=\exp\big(\sum_{\mathbf{m}}\theta^{\mathbf{m}}\sigma_{\mathbf{m}}^{[3]}\big),\label{eqexp}\end{equation}
 where $\theta^{\mathbf{m}}$ are real parameters. Because the state
is normalized, parameter $\theta^{\mathbf{0}}$ can be determined
by the other parameters $\{\theta^{\bar{\mathbf{m}}}\}$, which is
explicitly given by $\theta^{\mathbf{0}}=-\psi(\{\theta^{\bar{\mathbf{m}}}\})$,
where \begin{equation}
\psi(\{\theta^{\bar{\mathbf{m}}}\})=\mathrm{Tr}\big[\exp\big(\sum_{\bar{\mathbf{m}}}\theta^{\bar{\mathbf{m}}}\sigma_{\bar{\mathbf{m}}}^{[3]}\big)\big].\label{eqpsi}\end{equation}

The exponential form for a quantum state given by Eq. (\ref{eqexp})
ensures the positivity of the state automatically, which implies that
a bijective map can be built between the set of three-qubit states
with maximal rank and the set of $63$ real parameters $\{\theta^{\bar{\mathbf{m}}}\}$
via Eq. (\ref{eqexp}).

In addition, it is often instructive to regard the state in Eq. (\ref{eqexp})
as a thermal equilibrium state with the parameters $\{\theta^{\bar{\mathbf{m}}}\}$
describing the Hamiltonian of the three qubits. Then function $\psi(\{\theta^{\bar{\mathbf{m}}}\})$
in Eq. (\ref{eqpsi}) is the minus of the free energy in statistics
physics.

Another more widely used representation for the state $\rho^{[3]}$
is \begin{equation}
\rho^{[3]}(\{\eta^{\mathbf{m}}\})=\sum_{\mathbf{m}}\frac{\eta^{\mathbf{m}}}{8}\sigma_{\mathbf{m}}^{[3]},\label{eqm}\end{equation}
 where $\eta^{\mathbf{m}}$ are also real parameters. Here the normalization
of the state leads to $\eta^{\mathbf{0}}=1$, and the parameter $\eta^{\bar{\mathbf{m}}}$
is the average value of the Hermitian operator $\sigma_{\bar{\mathbf{m}}}^{[3]}$
for the state. Here we remark that not all real parameters $\{\eta^{\bar{\mathbf{m}}}\}$
will make Eq. (\ref{eqm}) semipositive, namely, being a legitimate
state.

\textit{Quantum relative entropy.} ---
Quantum relative entropy is a basic quantity in quantum information
theory \cite{Vedral,Wehrl}, which is defined by \begin{equation}
S(\rho||\sigma)=\mathrm{Tr}(\rho(\ln\rho-\ln\sigma)).\label{eqrelent}\end{equation}
 It satisfies the Klein's inequality \begin{equation}
S(\rho||\sigma)\ge0,\label{eqsemp}\end{equation}
 where the equality is satisfied if and only if $\rho=\sigma$. Quantum
relative entropy $S(\rho||\sigma)$ is a measure of distinguishability
for the state $\rho$ relative to the state $\sigma$. On one hand,
when the joint set between the support of the state $\rho$ and kernel
of the state $\sigma$ is not empty, the quantum relative entropy
becomes positive infinity, which means that we can definitely distinguish
the state $\rho$ from the state $\sigma$. On the other hand, the
Klein's inequality shows that only two identical states are completely
indistinguishable.

Quantum relative entropy has a desirable distance-like property: it
does not increase when part of the system is ignored, that is, for
a composite system $AB$, \begin{equation}
S(\rho^{(AB)}||\sigma^{(AB)})\ge S(\rho^{(A)}||\sigma^{(A)}).\label{eqqremon}\end{equation}
 Eq. (\ref{eqqremon}) implies that any quantum operation acting on
two states simultaneously can not increase their quantum relative
entropy.

Using the two representations for a quantum state given by Eq. (\ref{eqexp})
and Eq. (\ref{eqm}), we obtain \begin{eqnarray}
S(\rho^{[3]}||\rho^{\prime[3]})=\phi(\{\eta^{\bar{\mathbf{m}}}\})+\psi(\{{\theta^{\prime}}^{\bar{\mathbf{m}}}\})-\sum_{\bar{\mathbf{m}}}\eta^{\bar{\mathbf{m}}}{\theta^{\prime}}^{\bar{\mathbf{m}}},\label{eqrepqre}\end{eqnarray}
 where the function $\phi$ is the minus of the von Neaumann entropy,
namely, \begin{equation}
\phi(\{\eta^{\bar{\mathbf{m}}}\})=-S(\rho^{[3]}(\{\eta^{\bar{\mathbf{m}}}\})).\label{eqphi}\end{equation}
 Taking $\rho^{[3]}={\rho^{\prime}}^{[3]}$ in Eq. (\ref{eqrepqre}),
we get \begin{equation}
\phi(\{\eta^{\bar{\mathbf{m}}}\})+\psi(\{{\theta}^{\bar{\mathbf{m}}}\})-\sum_{\bar{\mathbf{m}}}\eta^{\bar{\mathbf{m}}}{\theta}^{\bar{\mathbf{m}}}=0.\label{eqLT}\end{equation}
 Eq. (\ref{eqLT}) is the Legendre transformation, which gives the
variable transformation \begin{eqnarray}
\eta^{\bar{\mathbf{n}}} & = & \frac{\partial\psi(\{\theta^{\bar{\mathbf{m}}}\})}{\partial\theta^{\bar{\mathbf{n}}}},\label{eqLTeta}\\
\theta^{\bar{\mathbf{n}}} & = & \frac{\partial\phi(\{{\eta}^{\bar{\mathbf{m}}}\})}{\partial\eta^{\bar{\mathbf{n}}}}.\label{eqLTtheta}\end{eqnarray}

Following Eq. (\ref{eqrepqre}) and Eq. (\ref{eqLT}), we obtain a
useful relation \begin{eqnarray}
 &  & S(\rho^{[3]}||\rho^{\prime\prime[3]})-S(\rho^{[3]}||\rho^{\prime[3]})-S(\rho^{\prime[3]}||\rho^{\prime\prime[3]})\nonumber \\
 &  & =\sum_{\bar{\mathbf{m}}}(\eta^{\bar{\mathbf{m}}}-{\eta^{\prime}}^{\bar{\mathbf{m}}})({\theta^{\prime}}^{\bar{\mathbf{m}}}-{\theta^{\prime\prime}}^{\bar{\mathbf{m}}}).\label{eqqrelink}\end{eqnarray}

\textit{Irreducible multiparty correlations.} ---
In this section, we will first review the definitions of irreducible
multiparty correlations given in Refs. \cite{Linden,Zhou}. Next we
will present new definitions on irreducible multiparty correlation
by extending Amari's work \cite{Amari}. Then we will show that these
two definitions are equivalent.

We consider the irreducible multiparty correlations in a given three-qubit
state $\rho_{\star}^{[3]}$.

\textbf{Definition $\mathbf{1}$:} First we define two sets of three-qubit
states as \begin{eqnarray}
M_{2}(\rho_{\star}^{[3]}) & = & \{\rho^{[3]}(\{\eta^{\bar{\mathbf{m}}}\})|\eta^{\bar{\mathbf{m}}}=\eta_{\star}^{\bar{\mathbf{m}}},\;\forall N_{0}(\bar{\mathbf{m}})\ge1\},\label{eqset1}\\
M_{1}(\rho_{\star}^{[3]}) & = & \{\rho^{[3]}(\{\eta^{\bar{\mathbf{m}}}\})|\eta^{\bar{\mathbf{m}}}=\eta_{\star}^{\bar{\mathbf{m}}},\;\forall N_{0}(\bar{\mathbf{m}})\ge2\}.\label{eqset2}\end{eqnarray}
 Obviously, the set $M_{2}(\rho_{\star}^{[3]})$ ($M_{1}(\rho_{\star}^{[3]})$)
is the set of three-qubit states that have the same two-qubit (one-qubit)
reduced density matrices as those of the state $\rho_{\star}^{[3]}$.
Next we will figure out the states with maximum von Neaumann entropy
in the above two sets respectively, which are defined by \begin{eqnarray}
\rho_{\star2}^{[3]} & = & {\arg\max}_{\rho^{[3]}\in M_{2}(\rho_{\star}^{[3]})}S(\rho^{[3]}),\label{eqmaxent2}\\
\rho_{\star1}^{[3]} & = & {\arg\max}_{\rho^{[3]}\in M_{1}(\rho_{\star}^{[3]})}S(\rho^{[3]}).\label{eqmaxent1}\end{eqnarray}
 Then the degree of irreducible three-party correlation and the degree
of irreducible two-party correlation are defined by \begin{eqnarray}
C_{3}(\rho_{\star}^{[3]}) & = & S(\rho_{\star2}^{[3]})-S(\rho_{\star}^{[3]}),\label{eqdef3cor1}\\
C_{2}(\rho_{\star}^{[3]}) & = & S(\rho_{\star1}^{[3]})-S(\rho_{\star2}^{[3]}).\label{eqdef2cor1}\end{eqnarray}
 The degree of the total correlation is \begin{equation}
C_{T}(\rho_{\star}^{[3]})=S(\rho_{\star1}^{[3]})-S(\rho_{\star}^{[3]}).\label{eqtcor1}\end{equation}

\textbf{Definition $\mathbf{2}$:} We also introduce two sets of three-qubit
states defined by \begin{eqnarray}
E_{2} & = & \{\rho^{[3]}(\{\theta^{\bar{\mathbf{m}}}\})|\theta^{\bar{\mathbf{m}}}=0,\;\forall N_{0}(\bar{\mathbf{m}})<1\},\label{eqset3}\\
E_{1} & = & \{\rho^{[3]}(\{\theta^{\bar{\mathbf{m}}}\})|\theta^{\bar{\mathbf{m}}}=0,\;\forall N_{0}(\bar{\mathbf{m}})<2\}.\label{eqset4}\end{eqnarray}
 Next we will find out the most indistinguishable states relative
to the state $\rho_{\star}^{[3]}$ in the above two sets, namely,
\begin{eqnarray}
\rho_{\star II}^{[3]} & = & {\arg\min}_{\rho^{[3]}\in E_{2}}S(\rho_{\star}^{[3]}||\rho^{[3]}),\label{eqminstat1}\\
\rho_{\star I}^{[3]} & = & {\arg\min}_{\rho^{[3]}\in E_{1}}S(\rho_{\star}^{[3]}||\rho^{[3]}).\label{eqminstat2}\end{eqnarray}
 Then the degrees of irreducible three-party correlation and irreducible
two-party correlation are defined by \begin{eqnarray}
C_{3}^{\prime}(\rho_{\star}^{[3]}) & = & S(\rho_{\star}^{[3]}||\rho_{\star II}^{[3]}),\label{eqc3p}\\
C_{2}^{\prime}(\rho_{\star}^{[3]}) & = & S(\rho_{\star II}^{[3]}||\rho_{\star I}^{[3]}).\label{eqc2p}\end{eqnarray}
 The total correlation is given by \begin{equation}
C_{T}^{\prime}(\rho_{\star}^{[3]})=S(\rho_{\star}^{[3]}||\rho_{\star I}^{[3]}).\label{eqctp}\end{equation}

The similar structure of the above two approaches motivates us to
consider whether they are equivalent or not. Actually, we have the
following theorem.

\textbf{Theorem: Definition $\mathbf{1}$} and \textbf{Definition
$\mathbf{2}$} on irrreducible multiparty correlations are equivalent.

\textbf{Proof:} In Tehorem $1$ of Ref. \cite{Zhou}, we have proved
that $\rho_{*2}^{[3]}\in E_{2}$ and $\rho_{\star1}^{[3]}\in E_{1}$.
Using Eq. (\ref{eqqrelink}), we have, $\forall\sigma_{2}^{[3]}\in E_{2}$
and $\forall\sigma_{1}^{[3]}\in E_{1}$, \begin{eqnarray}
S(\rho_{\star}^{[3]}||\sigma_{2}^{[3]}) & = & S(\rho_{\star}^{[3]}||\rho_{\star2}^{[3]})+S(\rho_{\star2}^{[3]}||\sigma_{2}^{[3]}),\label{eq2II}\\
S(\rho_{\star}^{[3]}||\sigma_{1}^{[3]}) & = & S(\rho_{\star}^{[3]}||\rho_{\star1}^{[3]})+S(\rho_{\star1}^{[3]}||\sigma_{1}^{[3]}).\label{eq1I}\end{eqnarray}
 Because of the Klein inequality (\ref{eqsemp}), Eq. (\ref{eq2II})
and Eq. (\ref{eq1I}) show that the state $\rho_{\star II}^{[3]}$
and the state $\rho_{\star I}^{[3]}$ are unique, and we have \begin{eqnarray}
\rho_{\star II}^{[3]} & = & \rho_{\star2}^{[3]},\label{eqII2}\\
\rho_{\star I}^{[3]} & = & \rho_{\star1}^{[3]}.\label{I1}\end{eqnarray}
 Inserting $\sigma_{2}^{[3]}=\sigma_{1}^{[3]}=\frac{\sigma_{\mathbf{0}}^{[3]}}{8}$
into Eq. (\ref{eq2II}) and Eq. (\ref{eq1I}), we obtain \begin{eqnarray}
C_{3}(\rho_{\star}^{[3]}) & = & C_{3}^{\prime}(\rho_{\star}^{[3]}),\label{eqc3eq}\\
C_{T}(\rho_{\star}^{[3]}) & = & C_{T}^{\prime}(\rho_{\star}^{[3]}).\label{eqcteq}\end{eqnarray}
 Inserting $\sigma_{2}^{[3]}=\rho_{\star1}^{[3]}$ into Eq. (\ref{eq2II}),
we find \begin{equation}
C_{2}(\rho_{\star}^{[3]})=C_{2}^{\prime}(\rho_{\star}^{[3]}).\label{eqc2eq}\end{equation}

In the process of the proof, we have shown that the states $\rho_{*2}^{[3]}$
and $\rho_{*1}^{[3]}$ are unique for a given three-qubit state $\rho_{*}^{[3]}$,
which is not easy to prove directly from Definition 1.

\textit{Properties.} ---
So far we have two equivalent definitions for the degrees of irreducible
multiparty correlations. However, we still lack the answer to the
fundamental question: Do the degrees of irreducible multiparty correlations
satisfy the basic requirements of a legitimate correlation measure?

Let us begin with a brief review of the basic requirements for a legitimate
correlation measure \cite{Henderson,Zhou1}. First, a correlation
measure is semipositive; Second, a correlation measure is invariant
under local unitary transformations; Third, a measure for a specific
type of correlation is zero if and only if the state has not such
type of correlation; Fourth, a correlation measure does not increase
under local operations .

According to the definitions of $C_{2}$, $C_{3}$, and $C_{T}$,
obviously they are semipositive. On one hand, a local unitary transformation
only changes the local basis, thus a bijective map can be built between
a state and the corresponding transformed state. On the other hand,
the involved functions, the von Neaumann entropy and the quantum relative
entropy, are invariant under local unitary transformations. Therefore,
the correlation measures $C_{2}$, $C_{3}$, and $C_{T}$ are invariant
under local unitary transformations.

The third and the fourth requirements can be analyzed conveniently
by the second definitions. In fact, $E_{2}$ is the set of the states
without irreducible three-party correlations, and $E_{1}$ is the
set of the states without correlations. According to the definitions
of $C_{3}$ ($C_{T}$), $C_{3}$ ($E_{T}$) of a state is zero if
and only if the state is in the set of $E_{2}$ ($E_{1}$). Notice
that we have not, and we may not need, an independent set of states
without irreducible two-party correlations.

Because of the monotonicity of quantum relative entropy (\ref{eqqremon}),
the necessary and sufficient condition for the measure $C_{T}$ ($C_{3}$)
satisfying the fourth requirement is that the set $E_{1}$ ($E_{2}$)
is closed under local quantum operations. Because any state in $E_{1}$
is a product state of the three parties, the set $E_{1}$ is indeed
closed under local quantum operations. Therefore, the total correlation
$C_{T}$ does not increase under local quantum operations.

However, we find that the set $E_{2}$ is not closed under local quantum
operations. This is explicitly demonstrated by the following counterexample.

\textbf{An counterexample:} The initial three-qubit state $\rho_{i}^{[3]}$
we consider is given by \begin{eqnarray}
\theta_{i}^{330}=1,\quad\theta_{i}^{303}=\theta_{i}^{001}=\frac{1}{\sqrt{2}},\end{eqnarray}
 where we only list the nonzero elements in the set $\{\theta^{\bar{\mathbf{m}}}\}$.
Obviously this state $\rho_{i}^{[3]}\in E_{2}$. To implement an local
operation on qubit $1$, we introduce an auxiliary qubit $a$, whose
initial state is given by $\theta_{a}^{3}=1$. Then we make the CNOT
operation with qubit $a$ the control qubit and qubit $1$ the target
qubit. The final state $\rho_{f}^{[3]}$ of the three qubits is given
by \begin{eqnarray}
 &  & \eta^{001}=\frac{\tanh(1)}{\sqrt{2}},\quad\eta^{033}=\eta^{303}=\frac{\tanh^{2}(1)}{\sqrt{2}},\nonumber \\
 &  & \eta^{330}=\tanh^{2}(1),\quad\eta^{331}=\frac{\tanh^{3}(1)}{\sqrt{2}}.\end{eqnarray}
 In the exponential form, the state $\rho_{f}^{[3]}$ can be written
numerically as \begin{eqnarray}
 &  & \theta^{001}\simeq0.650,\quad\theta^{033}=\theta^{303}\simeq0.336,\nonumber \\
 &  & \theta^{330}\simeq0.543,\quad\theta^{331}\simeq0.048.\end{eqnarray}
 Obviously, $\rho_{f}^{[3]}\notin E_{2}$.

The above counterexample shows that the irreducible three-party correlation
can be created by a local operation from a three-qubit state without
irreducible three-party correlations. Thus the measure $C_{3}$ can
increase under local quantum operations, which implies that, in this
sense, $C_{3}$ is not a legitimate correlation measure.

\textit{Discussions and summary.} ---
We formulate our results for a three-qubit state in the above discussions.
We emphasize that the formalism can be generalized to apply to any
multiparty state with finite dimensional Hilbert space. In the general
cases, the four Pauli matrices are taken place of by a complete set
of orthornamal Hermitian operators in the finite dimensional Hilbert
space \cite{Horn}. In addition, by using the exponential form, we
only deal with the multiparty quantum states with maximal rank. For
the multiparty quantum states without maximal rank, we can use the
continuity approach developed in Ref. \cite{Zhou} to investigate
the distribution of irreducible mulatiparty correlations in them.

Although our work is motivated by Amari's work on information geometry
\cite{Amari}, our presentaion focuses on the results but not on the
underlying mathematical structure. In fact, our presentation can be
understood without resorting to information geometry as follows. As
emphasized by Theorem $1$ in Ref. \cite{Zhou}, the exponential form
of a multiparty quantum state is of great significance in characterizing
irreducible multiparty correlations. More precisely, the theorem implies
that $E_{2}$ ($E_{1}$) is the set of three-qubit states without
irreducible three-party (two-party and three-party) correlations.
Thus Eqs. (\ref{eqc3p},\ref{eqc2p},\ref{eqctp}) can be introduced
via quantum relative entropy by the tradditional method widely adopted
in quantum information community \cite{Vedral,Plenio}.

According to the definitions of irreducible multiparty correlations,
the thermal equillibrium state of a Hamititonian with two-body interactions
will at most show irreducible two-party correlations and can not show
higher order irreducible multiparty correlations. However, we know
that a Hamitonian with two-body interactions can demonstrate topological
order \cite{Kitaev}, which is a type of irreducible macroscopic party-correlations.
How to concile this obvious paradox? This will be due to the degeneracy
of the ground states and the mechanism of spontaneous symmetry broken,
which makes the equibrium state is not simply written as the exponential
form of the Hamiltonian. This explains why a multiparty system with
short range interactions showing topological order always have degeneracy
of ground states.

In summary, we present the quantum version of Amari's work, and prove
that it is equivalent to the irreducible multiparty correlation defined
in \cite{Linden,Schneidman,Zhou}. Based on this new presentation,
we find a counterintuitive result: the irreducible three-party correlation
can be created by local operations from a three-party state with only
irreducible two-party correlations. This implies that the degree of
irreducible multiparty correlation can not be regarded as a legitimate
correlation measure independly. Local operations can not only destroy
high order correlations into lower order correlations, but also can
transform lower order correlations into higher order correlations.
We hope that this investigation will help us to characterize the multiparty
correlation structure in a multiparty quantum state, and finally improve
our understanding of the physics on multiparty correlations contained
in a multiparty quantum state.

This work is supported by NSF of China under Grant No. 10775176, and
NKBRSF of China under Grants No. 2006CB921206 and No. 2006AA06Z104.

\end{document}